\documentclass[journal,twoside,web]{ieeecolor}

\usepackage{etoolbox}
\makeatletter
\@ifundefined{color@begingroup}%
{\let\color@begingroup\relax
\let\color@endgroup\relax}{}%
\def\fix@ieeecolor@hbox#1{%
\hbox{\color@begingroup#1\color@endgroup}}
\patchcmd\@makecaption{\hbox}{\fix@ieeecolor@hbox}{}{\FAILED}
\patchcmd\@makecaption{\hbox}{\fix@ieeecolor@hbox}{}{\FAILED}

\usepackage{generic}
\usepackage{cite}
\usepackage{amsmath,amssymb,amsfonts}
\usepackage{algorithmic}
\usepackage{graphicx}
\usepackage{textcomp}
\usepackage{multirow}
\usepackage{multicol}
\makeatletter
\let\NAT@parse\undefined
\makeatother
\usepackage[colorlinks,citecolor=blue,urlcolor=blue,bookmarks=false,hypertexnames=true]{hyperref} 
\usepackage{soul}
\def\BibTeX{{\rm B\kern-.05em{\sc i\kern-.025em b}\kern-.08em
    T\kern-.1667em\lower.7ex\hbox{E}\kern-.125emX}}
\begin{document}
\title{Explicit Abnormality Extraction for Unsupervised Motion Artifact Reduction in Magnetic Resonance Imaging}
\author{Yusheng Zhou, Hao Li, Jianan Liu, Zhengmin Kong, Tao Huang, \IEEEmembership{Senior Member, IEEE}, Euijoon Ahn,\\ Zhihan Lv, \IEEEmembership{Senior Member, IEEE}, Jinman Kim, \IEEEmembership{Member, IEEE}, and David Dagan Feng, \IEEEmembership{Life Fellow, IEEE}
\thanks{This work was accepted by IEEE Journal of Biomedical and Health Informatics (DOI: 10.1109/JBHI.2024.3444771).}
\thanks{Yusheng~Zhou and Hao~Li contribute equally to the work and are co-first authors. Corresponding Author: Zhengmin~Kong}
\thanks{Yusheng~Zhou and Zhengmin~Kong are with the School of Electrical Engineering and Automation, Wuhan University, China. 
(Email: yushengzhou@whu.edu.cn; zmkong@whu.edu.cn)}
\thanks{Hao~Li is with the Department of Neuroradiology, University Hospital
Heidelberg, Heidelberg, Germany. (email: hao.li@med.uni-heidelberg.de)}
\thanks{Jianan~Liu is with Vitalent Consulting,
        Gothenburg, Sweden. (Email: jianan.liu@vitalent.se)}
\thanks{Tao~Huang and Euijoon Ahn are with the College of Science and Engineering,
        James Cook University, Cairns, Australia.
        (Email: tao.huang1@jcu.edu.au; euijoon.ahn@jcu.edu.au)}
\thanks{Zhihan Lv is with the
Department of Game Design, Faculty of Arts, Uppsala University, Sweden
(Email: lvzhihan@gmail.com)}
\thanks{Jinman Kim and David Dagan Feng are with the School of Computer Science, University of Sydney, Sydney, Australia (email: jinman.kim@sydney.edu.au; dagan.feng@sydney.edu.au).}
}

\maketitle

\begin{abstract}

Motion artifacts compromise the quality of magnetic resonance imaging (MRI) and pose challenges to achieving diagnostic outcomes and image-guided therapies. In recent years, supervised deep learning approaches have emerged as successful solutions for motion artifact reduction (MAR). One disadvantage of these methods is their dependency on acquiring paired sets of motion artifact-corrupted (MA-corrupted) and motion artifact-free (MA-free) MR images for training purposes. Obtaining such image pairs is difficult and therefore limits the application of supervised training. In this paper, we propose a novel UNsupervised Abnormality Extraction Network (UNAEN) to alleviate this problem. Our network is capable of working with unpaired MA-corrupted and MA-free images. It converts the MA-corrupted images to MA-reduced images by extracting abnormalities from the MA-corrupted images using a proposed artifact extractor, which intercepts the residual artifact maps from the MA-corrupted MR images explicitly, and a reconstructor to restore the original input from the MA-reduced images. The performance of UNAEN was assessed by experimenting with various publicly available MRI datasets and comparing them with state-of-the-art methods. The quantitative evaluation demonstrates the superiority of UNAEN over alternative MAR methods and visually exhibits fewer residual artifacts. Our results substantiate the potential of UNAEN as a promising solution applicable in real-world clinical environments, with the capability to enhance diagnostic accuracy and facilitate image-guided therapies. Our codes are publicly available at \url{https://github.com/YuSheng-Zhou/UNAEN}.

\end{abstract}


\section{Introduction}
\label{sec:introduction}

\IEEEPARstart{M}{agnetic} resonance imaging (MRI) is the most widely-used non-invasive medical imaging technique without radiation exposure. However, limited by the long acquisition time, e.g., 20-25 minutes for the whole brain, MRI is highly sensitive to the patient's movement \cite{artifact_explain}, and motion artifacts (MA) are frequently unavoidable. The MA is caused by an incorrect signal acquired and filled in the K-space, resulting in blurriness or ghosting artifacts. MA in MRI poses a substantial detriment to the entire image quality, impeding accurate interpretation of diagnostic information and hindering the efficacy of image-guided therapeutic interventions \cite{artifact_explain}. To tackle this problem, various methods have been proposed to prevent the patient movement and/or correct for MA \cite{restrain_motion, compressed_sensing, parallel_imaging, prospective_2, prospective_3, retrospective_1, retrospective_2}. Methods to prevent MA involves additional tools such as MR navigators \cite{MR_navigators_1, MR_navigators_2}, or external tracking devices \cite{external_tracking_devices}. Another approach is online motion measurement based on, e.g., the extended Kalman filter algorithm or orientation short tracking pulse-sequence, thereby compensating or reacquiring the K-space data partially in the case of extreme motion \cite{prospective_2, prospective_3} in a prospective manner. Although MA may be mitigated with these methods, they have not been widely used due to prolonged scan time and additional costs. Besides, statistical signal processing-based artifact correction is commonly utilized because of no prolonged scan time. Atkinson \emph{et al.} \cite{retrospective_1} applied an entropy-based criterion to an MR image to correct motion-induced artifacts. Batchelor \emph{et al.} \cite{retrospective_2} proposed a general matrix that mapped the transformation from the artifact-free image to the corrupted image, and then used the inversed mapping to reconstruct the artifact-free image. These techniques used prior statistical information to explore the influence of motion on the K-space data and build models to reconstruct motion artifact-free (MA-free) images. However, these methods still have limited capabilities due to the complexity and unpredictability of patients’ movements\cite{retrospective_1,retrospective_2}.

In recent studies, supervised deep learning techniques have been proposed for motion artifact reduction (MAR) in MRI, offering advantages such as avoiding additional acquisition equipment or prolonged scan time and achieving better performance \cite{deep_artifact_reduce1, deep_artifact_reduce2, deep_artifact_reduce3}. However, these methods heavily rely on a large number of training samples, where motion MA-free images serve as the reference ground truth to discern the disparities between motion artifact-corrupted (MA-corrupted) and MA-free images. These approaches have demonstrated superior performance compared to traditional statistical signal processing methods. However, the MA-corrupted images used in these studies are synthetically generated, and the acquisition of authentic paired MA-corrupted and MA-free images within clinical settings is exceedingly rare \cite{unpair_mar}, similar to the challenges encountered in acquiring paired high-resolution and low-resolution image pair in the super-resolution task \cite{MRI_super_resolution_no_paired_images_1,MRI_super_resolution_no_paired_images_2,BSR}. In clinical practice, the acquisition of images typically involves repeating measurements only when MA is evident. However, it should be noted that repeating the measurement does not guarantee the acquisition of MA-free images. Consequently, the availability of paired MA-corrupted and MA-free image sets is limited. Furthermore, patient movement during the imaging process can result in misalignment between images acquired from different series. These misaligned images are unsuitable for training, as their utilization can lead to highly blurred reconstructed images \cite{MRI_super_resolution_no_paired_images_2, BSR}. Notably, existing image registration algorithms currently available cannot effectively rectify such misalignments \cite{MRI_super_resolution_no_paired_images_2,ioct}. 

To overcome the issue of limited paired training data, unsupervised learning methods have been leveraged as a possible solution as they can be trained without paired images.  The well-recognized unsupervised learning in various computer vision tasks provides us with possible solutions to the aforementioned problems \cite{GAN, VAE, autoregressive, normalizing_flows}. Different from supervised methods, unsupervised learning can find hidden patterns or features from data without requiring feedback information from the ground truth and does not rely heavily on prior knowledge of the dataset. Specifically, several recent unsupervised learning methods have been implemented in medical image processing and shown promising results with unpaired training data on tasks that are similar to the MAR, such as ISCL \cite{ISCL} and UIDnet \cite{UIDnet} for medical image denoising in MRI and CT, ADN proposed by Liao \emph{et al.} \cite{ADN} for computed tomography (CT) metal artifact reduction, and CycleGAN \cite{CycleGAN} proposed by Zhu \emph{et al.} for understanding images style transfer. These methods use the mechanism of domain transfer, where they implicitly convert the images from one domain to another domain. In an implicit conversion, the network learns the pattern of MA in the source domain and converts the feature of the target domain to approach the source domain. The specific abnormal patterns in the target domain are not tackled and thus the network could be distracted by the other non-critical features.

As opposed to implicit methods, the proposed explicit method guided the network to focus on the MA pattern and extract the MA pattern from the MA-corrupted images, resulting in a more powerful representation learning ability of motion artifacts. Several previous studies have preferred to extract essential patterns explicitly. Tamada \emph{et al.} \cite{K-space_simulation_1} utilized a neural network to estimate the artifacts from the input images. Rai \emph{et al.} \cite{explicit_denoising} used the dictionary learning-based and residual learning-based methods to explicitly extract the noise from MRI/CT patches, and then preserved the noise characteristics by averaging them. Xiao \emph{et al.} \cite{SRNR} achieved 3D MRI super-resolution via directly learning the residual volume between the input and target using a modified U-Net. As a similar task to these studies, a motion artifact reduction network with explicit artifact extraction can be expected to achieve superior performance than the implicit methods.

To address the problems mentioned above, we proposed an unsupervised MRI artifact reduction framework with explicit artifact extraction. Specifically, given its property, we regarded the motion artifacts in MRI as a separable abnormality independent to the image content, where a network was applied to directly learn its representation from the MA-corrupted image, explicitly extracting the artifact residues. And then motion artifact reduction was achieved by simply subtracting the extracted artifacts from MA-corrupted images. The model equips with a cycle consistency and is trained with unpaired MA-free and MA-corrupted MR images and produces high-quality MA-corrected MR images. The contributions of this work are summarized as follows:
\begin{enumerate}
\item[$\bullet$] We proposed an unsupervised abnormality extraction network (UNAEN) to extract and remove MA by learning deep feature differences between unpaired MA-free images and MA-corrupted images, which are impractical to obtain in the real-world clinical environment.
\item[$\bullet$] Different from the existing methods, UNAEN aimed to explicitly extract abnormal MA information for improving the model's representation ability of motion artifacts, and corrected the abnormal information from the images. As a result, the feature distribution of MA-reduced images approached that of the MA-free images.
\item[$\bullet$] Experimental results showed that compared with other state-of-the-art unsupervised methods, our method obtained improved performance and generated images with superior quality.
\end{enumerate}

\section{RELATED WORK}\label{sec related}

\subsection{Deep learning-based Motion Artifact Reduction}

Because of the great prosperity of deep learning in the field of computer vision, deep learning-based retrospective MAR schemes  (especially convolutional neural networks, CNN) have been widely investigated. The CNN model can be trained with the MA-corrupted images as input and the MA-free images acquired from the same patient as ground truth. Johnson \emph{et al.}, as one of the pioneers using deep learning for MA correction, applied the deep neural network (DNN) to reconstruct the MA-corrected MR image from the MA-corrupted k-space \cite{deep_artifact_reduce2}. Han \emph{et al.} proposed a denoising algorithm based on U-net for the streak artifacts removal, which is induced in the radially acquired images \cite{deep_artifact_reduce1}. Sommer \emph{et al.} utilized a CNN to the extracted MA-only image, which was obtained by subtracting the MA-free image from the MA-corrupted image, resulting in less deformation \cite{deep_artifact_reduce3}. However, in most instances, it is very difficult to obtain paired MRI datasets for training neural networks. Although several motion simulation algorithms have been designed to solve this problem by generating synthetic MA-corrupted images from MA-free images using certain predefined movement patterns \cite{image_domain_simulation,K-space_simulation_1,K-space_simulation_2}, such approaches may incur a domain gap between the real and synthetic MA-corrupted images. The possible domain gap would downgrade the performance of the network, which is trained on simulated data but applied to real data \cite{MRI_super_resolution_no_paired_images_1,MRI_super_resolution_no_paired_images_2,BSR}.

\begin{figure*}[h]
\centering
\includegraphics[scale=0.23]{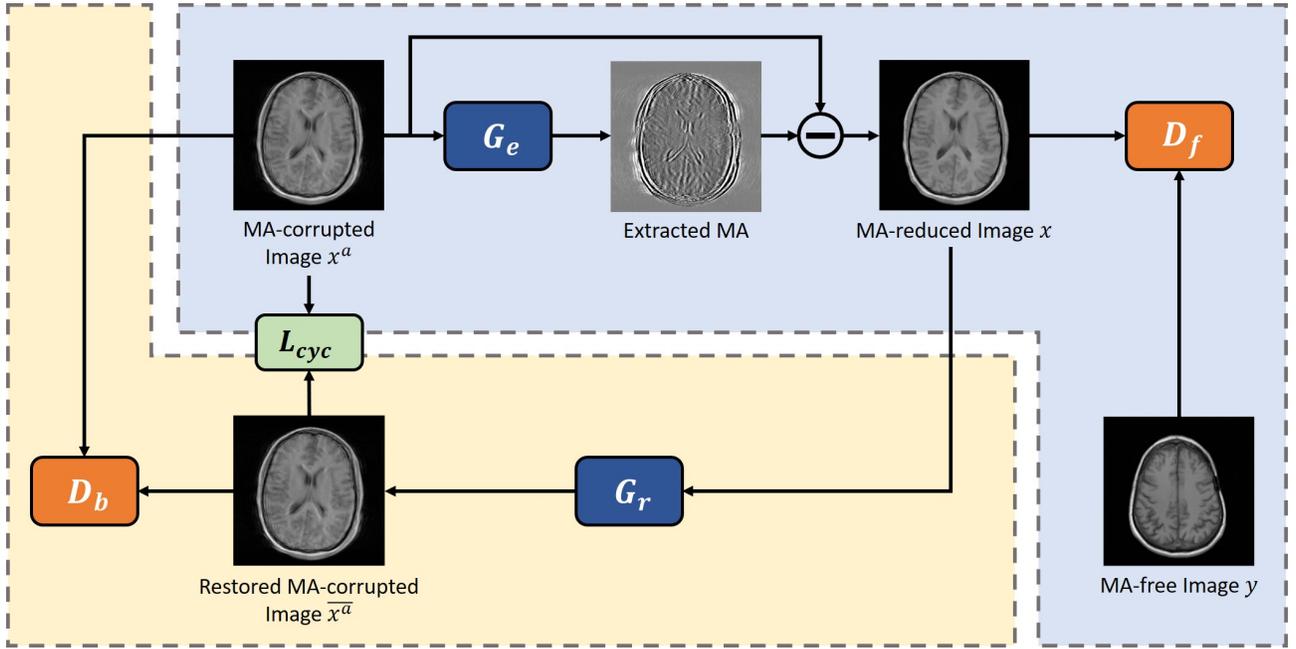}
\caption{The schematic of UNAEN. The proposed method contains a forward module (blue block), which reduces the artifacts of MA-corrupted images, and a backward module (yellow block), which restore the MA-corrupted images based on the forward module outputs. MA-reduced images can be obtained by explicitly extracting artifacts of MA-corrupted images using $G_e$ and deleting them directly by the forward module, where $D_f$ compares the MA-reduced images with the MA-free images to identify whether the artifact removal is successful. The backward module converts the MA-reduced images to the original input by $G_r$ and $D_b$ is used to check whether $G_r$ is restored successfully. There is a cycle consistency between the MA-corrupted images and the restored MA-corrupted images.}
\label{UNAEN}
\end{figure*}

\vspace{1cm}
\subsection{Unsupervised Image-to-Image Translation}

Most of the low-level computer vision tasks, e.g. denoising, super-resolution, MAR, etc., can be considered image-to-image translation, which converts images from one domain to another. In recent years, some training strategies based on unpaired images have attracted much attention. Deep Image Prior (DIP) \cite{DIP} demonstrated the randomly initialized network can generate a feasible hand-crafted prior for image denoising task. However, the disadvantage is the high consumption of resources for iterative computation for each image. Noise2Noise (N2N) \cite{N2N} and Noise2Void (N2V) \cite{N2V} only used noisy images to train a CNN denoiser. Although a satisfactory denoising effect can be achieved without noisy-clean image pairs, the global distribution of the noise is still required to choose the applicable loss functions. Recently, generative adversarial network (GAN) \cite{GAN} had shown great potential in image generation and representation learning, which was derived with many variations for different tasks. The GCBD \cite{GCBD} proposed by Chen \emph{et al.} illustrated that GAN can train to estimate the noise distribution of the noisy images. UIDnet \cite{UIDnet} applied a conditional GAN (cGAN) \cite{cGAN} to generate clean-pseudo noisy pairs for training a denoising network. CycleGAN \cite{CycleGAN, MedCycleGANv2} is a cyclic symmetric network consisting of two generators and two discriminators, which is mainly used for domain adaption. Cycle-MedGANv2\cite{MedCycleGANv2} improved CycleGAN by introducing two new cyclic feature-based losses (the cycle-perceptual loss and the cycle-style loss) to ensure cycle consistency and was applied to the rigid MR motion artifacts correction task. ISCL \cite{ISCL} added an extra network on the basis of CycleGAN to cooperate with the generators and estimate the noise distribution. By combining the generative model and disentanglement network, ADN \cite{ADN} constructed multiple encoders and decoders to separate the contents and artifacts in the CT images and get comparable results with supervised learning. As a common basis of the methods mentioned above, GAN is one of the most promising techniques at present to handle the distribution of complex data, whose studies have accumulated solid fundamental knowledge.

\section{PROPOSED METHOD}\label{sec methods}

\begin{figure*}[h]
\centerline{\includegraphics[scale=0.20]{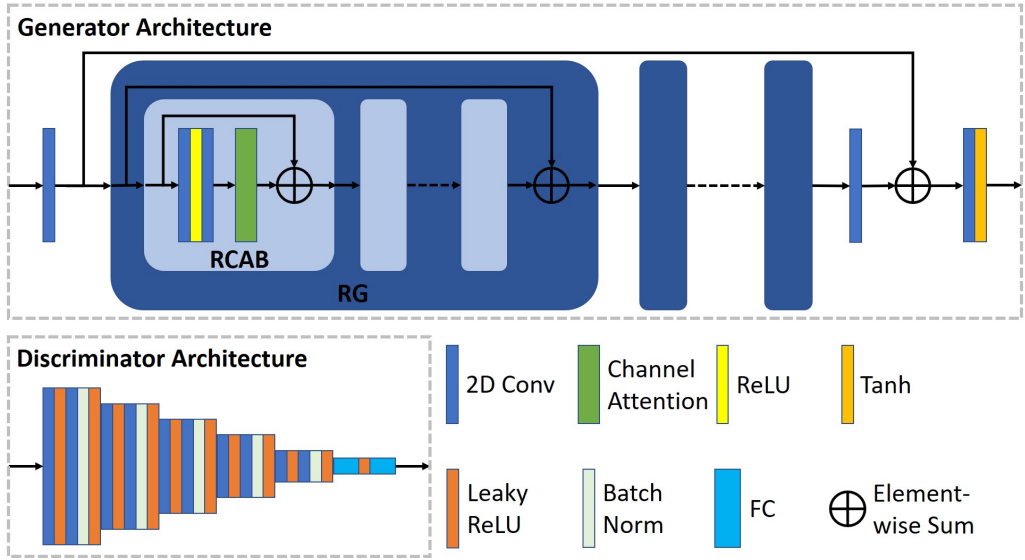}}
\caption{The detailed structures of generator and discriminator. The generator adopts the RCAN backbone with a depth of 5 residual groups (RG) and a long skip connection, and the discriminator is a VGG network.}
\label{network}
\end{figure*}

\subsection{Network Architecture}

Inspired by the cycle consistency of CycleGAN \cite{CycleGAN}, the UNAEN framework contains two modules: a forward module for artifact reduction and a backward module for artifact restoration as shown in Fig.\ref{UNAEN}. The forward module incorporates an artifact extractor, denoted as $G_e$, which is responsible for learning the artifact distribution within the MA-corrupted MR images. In parallel, the backward module employs an artifact reconstructor, denoted as $G_r$, to restore the corresponding original input based on the output generated by the forward module. The $G_e$ and $G_r$ are both generators of UNAEN. To train the generators, we employ $D_f$ and $D_b$ as discriminators in the forward and backward modules to distinguish between an MA-corrupted image and an MA-free image.

In the training process, unpaired images $\left\{ (x^a, y) | x^a\in X^a, y\in Y \right\}$ are used, where $X^a$ and $Y$ represent the MA-corrupted image patch and MA-free image patch, respectively. The MA-corrupted MR image $x^a$ is fed into $G_e$ to extract the artifact map $G_e(x^a)$, which affected the texture information of the images. The forward module generates the corresponding MA-reduced image $x$ by subtracting $G_e(x^a)$ from $x^a$:
\begin{equation}
\label{artifact reduction output}
x=x^a-G_e(x^a).
\end{equation}

$G_r$ is used to restore the generated $x$ and output the restored MA-corrupted image $x^a$, ensuring the forward module to translate an instance $x^a$ into a counterpart $x$ rather than any instance: 
\begin{equation}
\label{reconstruction result}
\overline{x^a}=G_r(x).
\end{equation}

There is a cycle consistency between $x^a$ and $\overline{x^a}$ and they are expected to be identical. Since $x$ and $y$ are unpaired and only have similar content, a forward discriminator $D_f$ is applied to identify the generated image $x$ and the real MA-free image $y$. To promote the reconstruction ability of $\overline{x^a}$, we train a backward discriminator $D_b$ to distinguish between the original input $x^a$ and restored MA-corrupted image $\overline{x^a}$. Therefore, the generators aim to generate samples that approximate the real data, while discriminators are not deceived by the output of the generators.

In this proposed framework, the generators and discriminators need to be trained alternately. In the inference step, only the trained $G_e$ is required. The MA-reduced images are obtained as long as the residual artifact maps are extracted by the $G_e$ from the corresponding MA-corrupted inputs.

The structures of generators and discriminators are shown in Fig.\ref{network}. The backbone of the generator is built by the Residual Channel Attention Network (RCAN) \cite{RCAN_1, RCAN_2} with a depth of 5 residual groups (RG). And each RG has 5 residual channel attention blocks (RCAB). We set the number of feature channels to 64. It is worth mentioning that the long-term connection existing between MA-corrupted image and extract MA map is not included in the structure of generator $G_e$, which is different from the general residual learning technique, since residual learning focuses on the missing information of source image compared to target image and supplementing while UNAEN explicitly learns the redundant artifact components in the MA-corrupted image and then subtracts it. Hence, $G_e$ simply learns artifact representations. The discriminators are built with convolutional units, each unit consists of a 3$\times$3 convolutional layer and a leaky rectified linear unit (leaky ReLU) activation layer \cite{leaky_ReLU}. The size of the feature map is reduced by half after every two convolution layers. All but the first unit have a batch normalization layer \cite{BN}. Similarly, the number of feature channels is set to 64 in the first convolutional layer of the discriminator and doubled after every two convolutional layers.

\subsection{Loss Functions} \label{loss functions}

In this design, three loss functions are selected and used in image restoration, which includes the L1 loss, Structural Similarity Index Measure (SSIM) loss \cite{ssim_1, ssim_2}, and adversarial loss in the training:
\begin{equation}
\label{L1 loss}
L_1(x,y)=\frac{1}{N} \sum_{i=1}^{N} \left | x-y \right |,
\end{equation}

\begin{equation}
\label{ssim loss}
L_{SSIM}(x,y)=\frac{1}{N} \sum_{i=1}^{N} \left | 1-SSIM(x,y)^{2}  \right |,
\end{equation}

\begin{equation}
\label{adv loss}
L_{adv}(x,D)=\frac{1}{N} \sum_{i=1}^{N} \sqrt{(D(x)-1)^{2}},
\end{equation}
where $D$ can be $D_f$ or $D_b$. SSIM is an indicator to quantify the similarity between two digital images. Eq.(\ref{SSIM}) shows the calculation of SSIM. In addition, we use the least square loss \cite{least_square_loss} as the adversarial loss in our model instead of the negative log-likelihood \cite{GAN} for stabilizing the training procedure.

To train $G_e$, a discriminator $D_f$ is used to classify the MA-reduced output $x$ as an MA-free image. The adversarial loss function $L_{G_e}$ is as follow:
\begin{equation}
\label{L_{Ge_adv}}
L_{G_e\_adv}(x,D_f)=\frac{1}{N} \sum_{i=1}^{N} \sqrt{(D_f(x)-1)^{2}}.
\end{equation}

To train $G_r$, we use a discriminator $D_b$, which classifies the restored MA-corrupted result $\overline{x^a}$ as the original MA-corrupted image. The following adversarial loss function is used to train the $G_r$:
\begin{equation}
\label{L_{Gr_adv}}
L_{G_r\_adv}(\overline{x^a},D_b)=\frac{1}{N} \sum_{i=1}^{N} \sqrt{(D_b(\overline{x^a})-1)^{2}}.
\end{equation}

Moreover, we adopt the cycle consistency loss to restrain the restoration of $\overline{x^a}$. It is calculated as a weighted sum of L1 loss and SSIM loss between the inputs and reconstruction images:
\begin{equation}
\label{L_{Gr_cyc}}
L_{cyc}(x^a,\overline{x^a})=L_1(x^a,\overline{x^a})+\lambda _{SSIM}*L_{SSIM}(x^a,\overline{x^a}),
\end{equation}
where $\lambda _{SSIM}$ is the weight of SSIM loss. We set $\lambda _{SSIM}$ = 0.5 in our experiments.

The final objective function that optimizes the $G_e$ and $G_r$ networks can be represented as:
\begin{equation}
\label{L_G}
L_G=\lambda _{G_e\_adv}*L_{G_e\_adv}+\lambda _{G_r\_adv}*L_{G_r\_adv}+L_{cyc},
\end{equation}
where $\lambda _{G_e\_adv}$ and $\lambda _{G_r\_adv}$ are the weights of the adversarial losses of $G_e$ and $G_r$, respectively. Given that $L_{cyc}$ dominates consistent improvement in the network’s performance as a training supervisor and stabilize the training process, $\lambda _{G_e\_adv}$ and $\lambda _{G_r\_adv}$ were empirically set to 0.1 to balance the loss components, and achieve the most stable convergence and best performance in our experiments.

\subsection{Motion Simulation}
\label{subsec:Motion Simulation}

We adopt the  method proposed by Li \emph{et al.} \cite{motion_simulation} to simulate the motion in MR images. Splicing lines from multiple K-space is used to simulate the generation of real motion artifacts. Firstly, a group of images is generated from the original images by rotating them in specific directions and to specific degrees as done in \cite{motion_simulation}. The duration and frequency of motion for any movement pattern control the severity of the artifact. The original images and the generated images are transformed to K-space using FFT, and K-space segments of the original image are replaced with segments from the generated images’ K-spaces, according to predefined rotation directions and rotation degrees. Finally, the damaged original K-space data is transferred back to the image domain by iFFT to obtain the simulation MA- corrupted MR image. In the motion simulation process, we use the echo group (EG) as the minimum time period unit to obtain a certain number of successive echoes, and the duration of any action is an integer multiple of EG. To simulate the motion of the patient’s head, we set the original images to be rotated 5 degrees to the left and to the right in the plane. Specifically, we use the K-space segments of the rotated images to periodically replace the K-space segments of the original image from the center lines to the edge lines.

\section{EXPERIMENTS AND RESULTS}\label{sec results}

\subsection{Dataset Description}

In this study, the fastMRI brain dataset \cite{fastMRI_1, fastMRI_2} was used to evaluate our method. It includes 6970 fully sampled brain MRIs (3001 at 1.5T and 3969 at 3T) collected at NYU Langone Health on Siemens scanners using T1-weighted, T2-weighted, and FLAIR acquisitions. Some of the T1-weighted acquisitions included admissions of contrast agents. We randomly selected 5000 images from the T1 weighted slices with 3T field strength, the matrix size of the images is 320$\times$320. 

We also conduct another experiment on the BraTS dataset \cite{BraTS_1, BraTS_2, BraTS_3}, which consists of images from 369 participants with brain tumors. The contrast-enhanced T1w (T1CE) MR images from the BraTS dataset are chosen in this study. The T1CE images are acquired in the axial plane, the matrix size was 240$\times$240$\times$155 with an isotropic resolution of 1.0 mm. We apply the same data processing flow and configurations as the fastMRI to simulate motion artifacts. 

In our experiments, the slices without anatomical structures are discarded. All selected images are corrupted from the K-space by using the motion simulation algorithm (Section \ref{subsec:Motion Simulation}). Specifically, 1 EG contains 10 echos, and the movement interval $T_S$ is set to 3 EG, 6 EG, and 9 EG, resulting in a K-space corrupted line ratio of 75\%, 60\%, and 50\%, respectively. Then the dataset was divided into training, validation, and test sets. Among them, the fastMRI dataset was split based on images while the BraTS dataset based on patients. The unsupervised MAR method only requires unpaired MA-free MR images and MA-corrupted MR images, thus we further divide the training set into two groups. One group contains only MA-free images as a learning target while the other group contains only MA-corrupted images as input to the model. The validation set is used to monitor the networks’ performance during training and the test set to evaluate the networks after training. All of the images are normalized to the range of 0 to 1. To save computation resources, each image is cropped into 128$\times$128 patches. After cropping, the numbers of training patches, validation patches and test patches of fastMRI are 36000, 4500 and 4500 respectively, and the numbers of BraTS are 48000, 6000 and 6000 respectively.

\begin{figure*}[h]
\centerline{\includegraphics[scale=0.13]{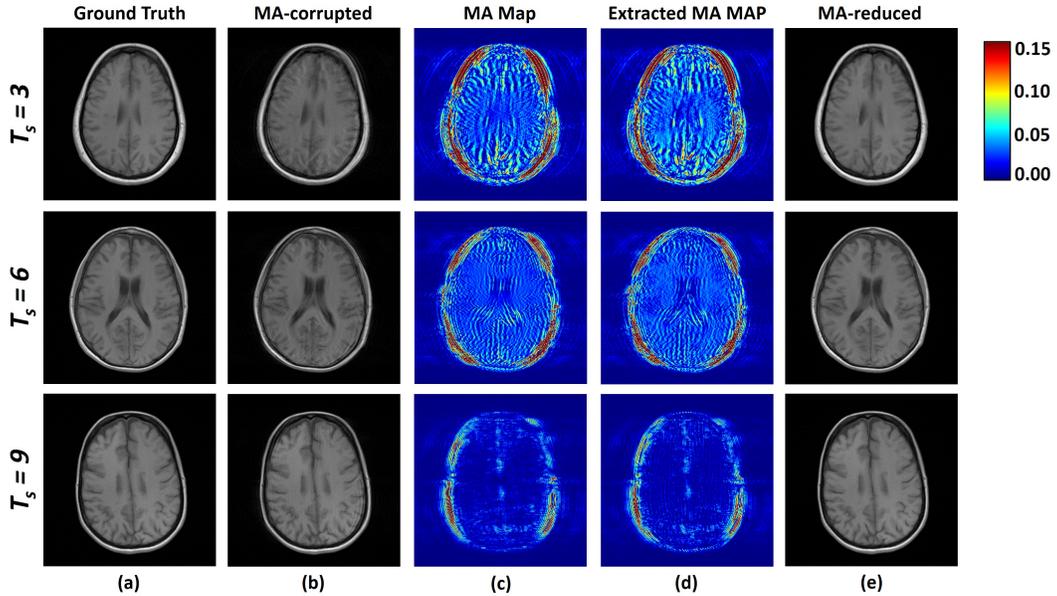}}
\caption{Illustration of motion artifact extractions with varying degrees of severity in each row. MA Maps column (c) denotes the error between MA-free images (Ground Truth in column (a)) and corresponding MA-corrupted images (column (b)). The proposed network was trained to extract residual artifact maps (Extracted MA MAP in column (d)) explicitly. The MA-reduced column (e) displays the restored images after MA reduction. Highly consistent patterns can be observed between the MA Map column and extracted MA Map column, revealing the effectiveness of the explicit MA extraction.}
\label{extraction}
\end{figure*}

\subsection{Evaluation Metrics}

In order to make a comprehensive comparison, we use SSIM and PSNR as the evaluation metrics in our experiments.

As mentioned in Section \ref{loss functions}, SSIM can quantify the similarity of two images. It is defined to compare the brightness, contrast, and structure between the MA-reduced output x and the ground truth. The SSIM is in the range of [-1, 1] and a larger value represents a better performance. The specific expression is as below:
\begin{equation}
\label{SSIM}
SSIM(X,Y)=\frac{(2\mu_X\mu_Y+C_1)(2\sigma_{XY}+C_2)}{(\mu^2_X+\mu^2_Y+C_1)(\sigma^2_X+\sigma^2_Y+C_2)},
\end{equation}
where $\mu$ and $\sigma$ donate the mean and standard deviation of the images, respectively ($\sigma^2_{XY}$ donates the covariance of x and y). $C_1$ and $C_2$ are constants.

The PSNR is another widely employed image quality indicator, which represents the ratio between the maximum possible signal value and the interference noise value that affects the signal representation accuracy. It is usually measured in decibels (DB) and a higher value indicates a lower distortion. PSNR can be calculated according to the following formula:
\begin{equation}
\label{PSNR}
PSNR=10\log_{10}{\frac{Max Value^2}{MSE}},
\end{equation}

\begin{equation}
\label{MSE}
MSE=\frac{1}{mn} \sum_{i=0}^{m-1}\sum_{j=0}^{n-1} [I(i,j)-K(i,j)]^2,
\end{equation}
where ${Max Value}$ is the largest possible pixel value and ${MSE}$ calculates the mean square error of two images. It is difficult for human eyes to perceive the difference when PSNR exceeds 30.

\subsection{Implementation Details}

All our experiments were implemented on a workstation with 64GB RAM and two NVIDIA GeForce RTX 2080 Ti graphics cards. Pytorch 1.8.1 was used as the back end. Before each epoch of the training process, all MA-free and MA-corrupted image patches were shuffled. We trained our model for 50 epochs using the ADAM optimizer with $\beta_1 = 0.9, \beta_2 = 0.99$, and set the batch size to 4. In each batch, the MA-free patches and the MA-corrupted patches fed to the networks were unpaired. The initial learning rate was set to 0.0001 and dropped by half in every 10 epochs. The generators were trained twice every time the discriminators were trained.

\begin{table}
    \setlength{\tabcolsep}{3pt}
    \centering
    \caption{Quantitative comparison with different combinations of generators and discriminators of UNAEN for MRI motion artifact reduction. \textbf{Bold} font represents the best.}
    \label{table1}
    \scalebox{1.2}{
    \begin{tabular}
    {|p{80pt}|p{30pt}|p{30pt}|}
    \hline
    Methods & SSIM & PSNR  \\
    \hline
    explicit w/ $G_r$ & \textbf{0.9126} & \textbf{30.5387} \\
    implicit w/ $G_r$ & 0.9087 & 30.4296 \\
    explicit w/o $G_r$ & 0.9086 & 29.8300 \\
    implicit w/o $G_r$ & 0.9057 & 29.5269 \\
    \hline
    \end{tabular}
    }
\end{table}

\begin{table*}
\setlength{\tabcolsep}{3pt}
\centering
\caption{Quantitative comparison with the state-of-the-art unsupervised networks for MRI motion artifact reduction. \textbf{Bold} font represents the best and \underline{underline} represents the second best.}
\label{table2}
\scalebox{1.2}{
\begin{tabular}{|p{30pt}|p{100pt}|p{25pt}|p{30pt}|p{25pt}|p{30pt}|p{25pt}|p{30pt}|}
\hline

\multirow{2}{*}{Dataset} & \multirow{2}{*}{Methods} & \multicolumn{2}{c|}{$T_S$=3 EG} & 
\multicolumn{2}{c|}{$T_S$=6 EG} & \multicolumn{2}{c|}{$T_S$=9 EG} \\
\cline{3-8}
& & SSIM   & PSNR  & SSIM   & PSNR  & SSIM   & PSNR  \\
\hline \hline

\multirow{5}{*}{fastMRI} & Before Reduction
& 0.7981 & 26.6165 & 0.8824 & 30.4109 & 0.9225 & 33.4192 \\
\cline{2-8}
& UIDnet \cite{UIDnet}
& 0.8551 & 27.1392 & 0.9168 & 30.4248 & 0.9411 & 32.5677 \\
& Cycle-MedGANv2 \cite{MedCycleGANv2}
& 0.8714 & 27.4449 & 0.9263 & 31.1473 & 0.9559 & 33.4017 \\
& ISCL \cite{ISCL}
& 0.8958 & 29.3085 & 0.9410 & 32.4944 & 0.9586 & 34.4717 \\
& DR-CycleGAN \cite{DR-CycleGAN} & \underline{0.9066} & \underline{30.4468} & \underline{0.9484} & \underline{33.2903} & \underline{0.9621} & \underline{34.7605} \\
& UNAEN (Ours)
& \textbf{0.9126} & \textbf{30.5387} & \textbf{0.9504} & \textbf{33.5448} & \textbf{0.9674} & \textbf{35.9265} \\
\hline \hline

\multirow{5}{*}{BraTS} & Before Reduction
& 0.7457 & 26.4940 & 0.8281 & 30.1854 & 0.8813 & 33.0922 \\
\cline{2-8}
& UIDnet \cite{UIDnet}
& 0.7663 & 26.3967 & 0.8376 & 26.1034 & 0.9116 & 30.6354 \\
& Cycle-MedGANv2 \cite{MedCycleGANv2}
& 0.8613 & 25.9863 & 0.8610 & 28.1681 & 0.9684 & 34.1405 \\
& ISCL \cite{ISCL}
& 0.8998 & \underline{27.9572} & \underline{0.9531} & \underline{32.1735} & \textbf{0.9719} & \textbf{34.7336} \\
& DR-CycleGAN \cite{DR-CycleGAN} & \underline{0.9091} & 27.7273 & 0.9277 & 29.8630 & 0.9498 & 31.0975 \\
& UNAEN (Ours)
& \textbf{0.9112} & \textbf{28.2383} & \textbf{0.9665} & \textbf{33.2732} & \underline{0.9704} & \underline{34.1600} \\
\hline

\end{tabular}
}
\end{table*}

\subsection{Illustration of Motion Artifact Extraction}

Figure \ref{extraction} depicts the motion artifact extractions with varying degrees of artifact severity. The (c) column shows the MA maps, which denote the error between the MA-free images in the (a) column and the corresponding MA-corrupted images in the (b) column. In the proposed method, we trained a residual channel attention network as the artifact extractor to explicitly extract residual artifact maps, which are visualized in the (d) column of Figure 3. The (e) column shows the restored images after MA reduction. The highly consistent patterns between the MA maps and extracted MA maps reveal the effectiveness of the explicit MA extraction. With a certain degree of consistency between the real MA and extracted MA, UNAEN achieved MRI MA reduction with comparable quality to MA-free images by simply subtracting the extracted MA from the MA-corrupted images.

\begin{figure*}[h]
\centerline{\includegraphics[scale=0.14]{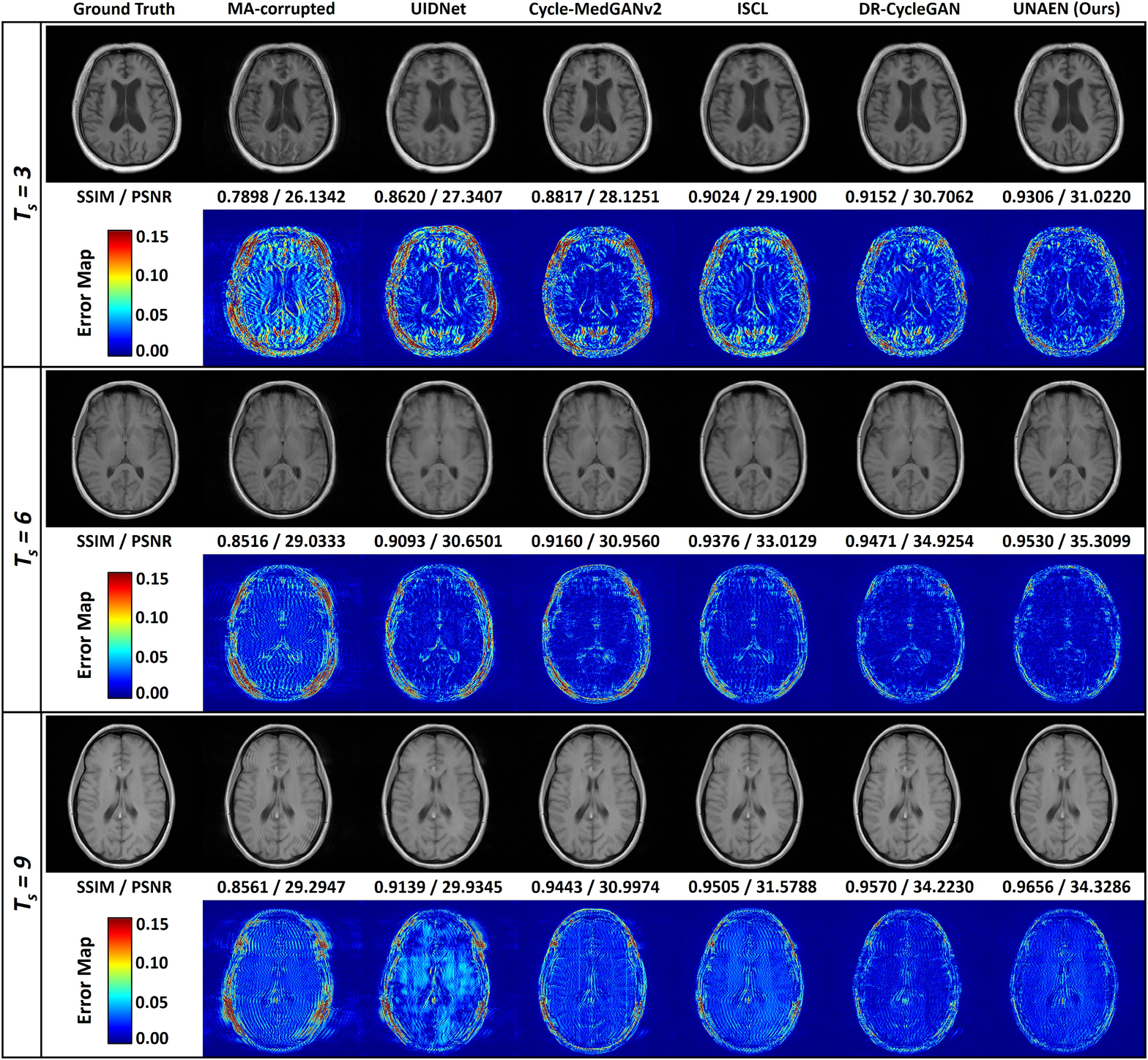}}
\caption{Comparison of the qualitative performance of UNAEN and other unsupervised models on the fastMRI brain dataset. There are visualizations of the artifact reduction results with varying degrees of artifact severity and corresponding error heat maps showing the difference between ground truth and each result.}
\label{fastMRI_result}
\end{figure*}

\begin{figure*}[h]
\centerline{\includegraphics[scale=0.19]{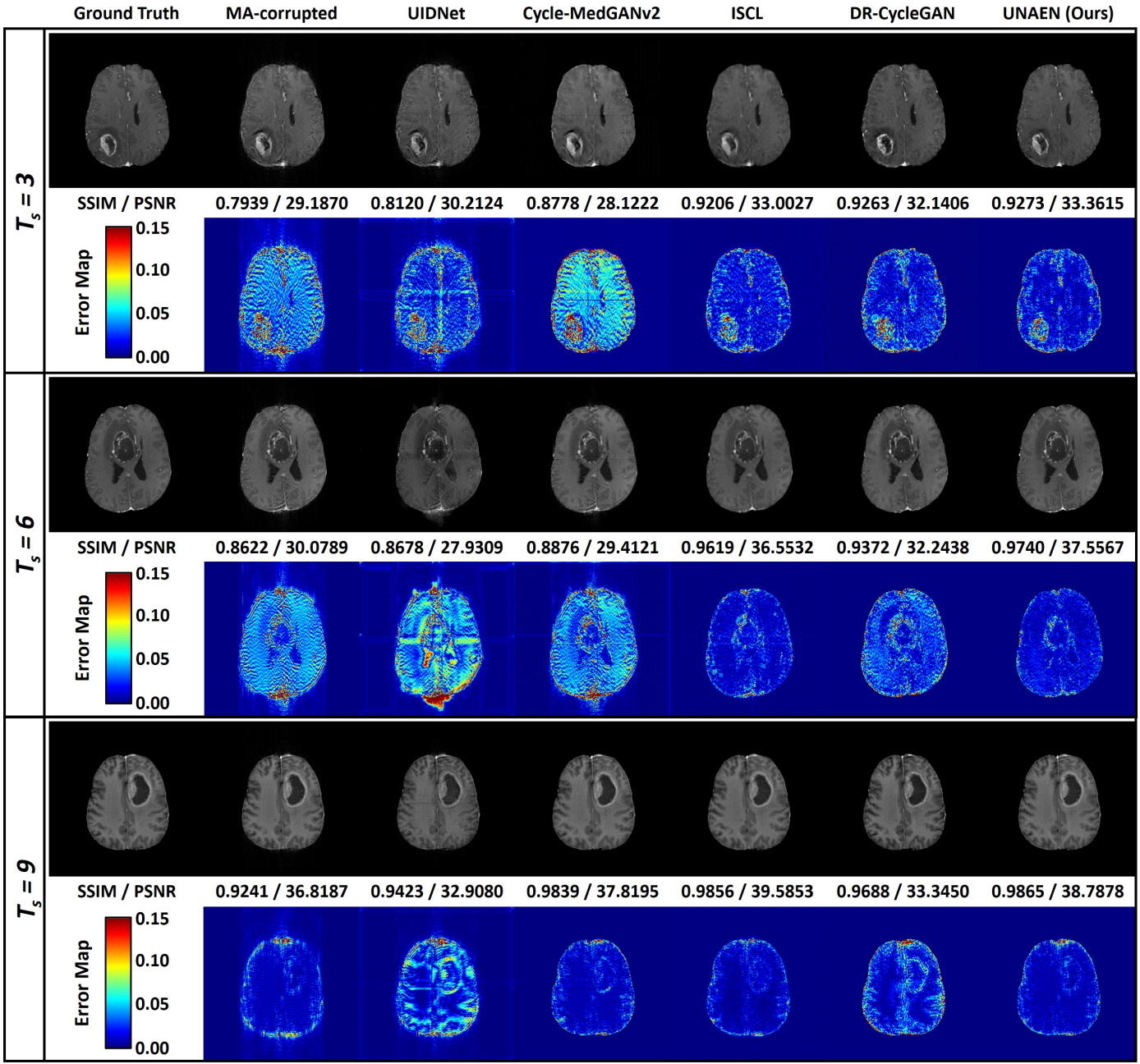}}
\caption{Comparison of the qualitative performance of UNAEN and other unsupervised models on the BraTS dataset. This visualized the artifact reduction results with varying degrees of artifact severity and corresponding error heat maps showing the difference between ground truth and each result.}
\label{BraTS_result}
\end{figure*}

\begin{figure*}[h]
\centerline{\includegraphics[scale=0.16]{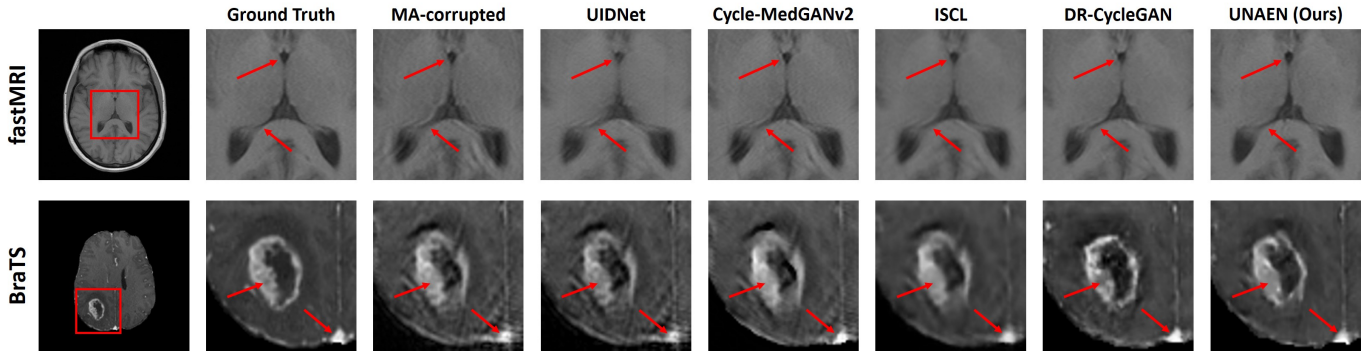}}
\caption{MA reduction details of UNAEN and other SOTA methods. The top row shows the partially enlarged images of MAR results from the fastMRI dataset with $T_s$=6 EG, and the bottom row on the BraTS dataset with $T_s$=3 EG. The arrows point out the small structures where UNAEN has sharper boundaries and fewer residual artifacts than other networks.}
\label{partial_detail}
\end{figure*}

\subsection{Ablation Study}

We verified the effectiveness of the explicit strategy and $G_r$ with various network configurations on the fastMRI brain dataset. As shown in Table \ref{table1}, the implicit methods refer to generating MA-reduced images directly by $G_e$ without extracting the MA pattern, whereas the methods with $G_r$ refer to constituting one-cycle consistency in the network. The results show that the metrics of the methods with $G_r$ are better than those of methods without $G_r$, where the SSIM / PSNR are up to 0.004 / 0.9027 dB higher, and the explicit methods outperform the implicit ones. Besides, solely activating the explicit strategy or the $G_r$ can raise the SSIM value to a comparable extent, and the latter can improve the PSNR value further. The combination of the explicit strategy and $G_r$ achieved the best results in the ablation experiments.

\subsection{Comparison with the State-of-the-art Networks}

Table \ref{table2} shows the comparison between the UNAEN and other SOTA models on two datasets with varying severity of MA. Lower SSIM and PSNR of the MA-corrupted images indicate higher severity of the MA.

The top half of Table \ref{table2} shows the experimental results on the fastMRI brain dataset. We observe that the proposed unsupervised model is significantly superior to all comparison unsupervised methods, where the SSIM and PSNR values are up to 0.0575 and 3.3995 dB higher than the other methods. Fig.\ref{fastMRI_result} visualizes the artifact reduction effects of different models and shows the qualitative performance on three degrees of artifact severity by displaying the reduction results and corresponding error heat maps compared to ground truth. All five unsupervised methods we compared (UIDnet, Cycle-MedGANv2, ISCL, DR-CycleGAN, and UNAEN) successfully reduce the motion artifact. UIDnet appears to have the weakest artifact reduction ability and its outputs still retain significant artifact traces in the marginal region of the tissue. Similarly, Cycle-MedGANv2 generates blurry images even though it has a higher SSIM and PSNR than UIDnet. ISCL and DR-CycleGAN have improved artifact reduction performance and image quality. However, evident errors on the boundaries of distinct soft tissues are observed in the reduction results, as shown in the error heat maps in Fig.\ref{fastMRI_result}. More details can be observed in Fig.\ref{partial_detail}. On the contrary, UNAEN achieves higher metrics values and minimizes errors, and with the increase in artifact severity, the performance gap with other methods is larger. In summary, UNAEN outperforms other compared models in terms of overall image quality and feature details in the experiment of the fastMRI brain dataset.

The experimental results on the BraTS dataset are shown in the bottom half of Table \ref{table2}. UNAEN achieves the best or second-best SSIM and PSNR, which are comparable to the ISCL and outperformed all of the other networks. As to the qualitative comparison visualized in Fig.\ref{BraTS_result} for the motion artifact reduction of the BraTS dataset, UIDnet and ISCL generate blurry images while UNAEN, DR-CycleGAN and Cycle-MedGANv2 generate clear ones. Specifically, DR-CycleGAN showed performance comparable to ISCL and close to UNAEN at the high artifact severity cases, but did not maintain an advantage in processing mild cases, which was just the opposite of Cycle-MedGANv2. Both ISCL and UNAEN have achieved high metrics in the BraTS dataset. It is observed that ISCL sometimes achieved higher metrics values with the BraTS dataset, particularly with mild motion artifact severity. However, the images restored by ISCL are blurry with the detained anatomical structures over-smoothened as shown in Fig.\ref{partial_detail}. Contrarily, the UNAEN-restored images show sharper boundaries between different tissues. This is also observed in the fastMRI dataset in the same figure. Therefore, the higher metrics values are not helpful for the clinical diagnosis from the perspective of visual effects.

Considering that the movement of patients does not exist only in a single plane in practice, we further carried out experiments of MRI inter-plane motion artifact reduction. Specifically, we randomly selected the same number of images from the BraTS dataset as that used in previous in-plane experiments. Then, we simulated artifacts in axial and sagittal planes with $T_S$=9 EG. The artifact simulation strategy is the same as described in Section \ref{subsec:Motion Simulation}. The experimental results conducted on the newly processed dataset were gathered in Table \ref{table3}. We can observe that Cycle-MedGANv2 obtains the lowest SSIM and DR-CycleGAN has the lowest PSNR, while UNAEN still maintained its advantage and achieved the highest artifact reduction effect among these methods. To sum up, UNAEN shows an overall superior performance for both in-plane and inter-plane MAR.

\begin{table}
    \setlength{\tabcolsep}{3pt}
    \centering
    \caption{Quantitative comparison with the state-of-the-art unsupervised networks for MRI inter-plane MAR. \textbf{Bold} font represents the best and \underline{underline} represents the second best.}
    \label{table3}
    \scalebox{1.2}{
    \begin{tabular}
    {|p{80pt}|p{30pt}|p{30pt}|}
    \hline
    Methods & SSIM & PSNR  \\
    \hline
    Before Reduction & 0.7439 & 32.1020 \\
    \hline
    UIDnet \cite{UIDnet} & 0.7698 & 31.4149 \\
    Cycle-MedGANv2 \cite{MedCycleGANv2} & 0.7637 & 30.4147 \\
    ISCL \cite{ISCL} & \underline{0.9045} & \underline{31.5749} \\
    DR-CycleGAN \cite{DR-CycleGAN} & 0.8942 & 28.8926 \\
    UNAEN (Ours) & \textbf{0.9107} & \textbf{31.6560} \\
    \hline
    \end{tabular}
    }
\end{table}

To further compare the performance, we then test the floating point operations (FLOPs), the number of network's parameters (Params) and the inference time (Time) of these comparison methods to demonstrate model efficiency, and the results are shown in Table \ref{table4}. In our case, the Time we measured is the average inference time of 200 images with a matrix size of 128$\times$128. We observe that UIDnet shows the superior efficiency with the second lowest FLOPs, the least Params and the fastest execution speed of 9.14G, 0.56M and 6.47ms respectively, while it performs the worst in motion artifact reduction. In contrast, by achieving the best motion artifact reduction performance, UNAEN shows relatively high FLOPs and inference time.

\begin{table}
    \setlength{\tabcolsep}{3pt}
    \centering
    \caption{Quantitative comparison with the state-of-the-art unsupervised networks for model efficiency.}
    \label{table4}
    \scalebox{1.2}{
    \begin{tabular}
    {|p{80pt}|p{30pt}|p{30pt}|p{30pt}|}
    \hline
    Methods & FLOPs & Params & Time \\
    \hline
    UIDnet \cite{UIDnet} & 9.14G & 0.56M & 6.47ms \\
    Cycle-MedGANv2 \cite{MedCycleGANv2} & 33.93G & 2.08M & 20.05ms \\
    ISCL \cite{ISCL} & 4.00G & 1.26M & 6.63ms \\
    DR-CycleGAN \cite{DR-CycleGAN} & 17.58G & 11.14M & 9.48ms \\
    UNAEN (Ours) & 33.93G & 2.08M & 19.67ms \\
    \hline
    \end{tabular}
    }
\end{table}

\section{DISCUSSION }\label{sec discussion}

Compared to other proposed methods that use implicit domain transfer approaches, UNAEN's explicit artifact extraction approach has several advantages. UIDnet trains a cGAN \cite{cGAN} which adds artifacts to clean images in order to generate paired images to train a de-artifacts network under supervision. Due to UIDnet's non-End-to-End training strategy, more errors will be induced than in other models, limiting the ability to remove artifacts and resulting in the lowest SSIM and PSNR in the experiments. Therefore, significant artifact traces are retained in the image, leading to inaccurate surgery or therapy doses.

As another unsupervised network for domain transfer tasks, Cycle-MedGANv2 can transfer images between different styles. To generate a tighter mapping space, two symmetric generators are used to realize the implicit conversion between the MA-corrupted and the MA-free image domains. However, the experimental results demonstrate that UNAEN outperformed Cycle-MedGANv2 with a big gap, revealing the effectiveness of explicit artifact extraction over the implicit domain transfer.

ISCL is a variation of CycleGAN that adds an additional extractor to cooperate with generators. The generators are responsible for direct conversion between image domains, while the extractor can extract the artifacts. The experimental results in Fig.\ref{partial_detail} and Table \ref{table2} showed that ISCL can further improve the SSIM and PSNR values at the cost of image blurriness. 

As another variation of CycleGAN, DR-CycleGAN, a network-intensive method, does not directly transfer images between different domains, but requires specific encoders to extract the artifact features and content features separately, and MA-reduced or original images can be restored by specific decoders through the combination of the extracted features. Although this strategy enhances the disentanglement of artifact and content of MRI, the disadvantages are obvious. The increase in the number of networks makes the training of the model more complicated, and the performance on the data with weak artifact severity or data with entangled artifacts will be significantly limited, as shown in our experimental results on BraTS dataset with $T_s$=6 EG and $T_s$=9 EG.

Different from the Cycle-MedGANv2, ISCL and DR-CycleGAN, UNAEN only adopts one cycle consistency and is more stable in training than Cycle-MedGANv2 and generates clearer results than ISCL. The abandonment of redundant training makes the model pay more attention to the artifact removal process, while explicit extraction strategy fits the artifacts reduction task and promotes the representation ability of artifacts. Experimental results demonstrate that our modifications can successfully extract the residual artifacts from the MA-corrupted images and suppress the motion artifact with significantly improved metrics values and enhanced quality of MA-reduced images.

UNAEN shows promising potential in correcting MA to avoid the misrepresentation of anatomical structures in the images. As a result, UNAEN can reduce the artifacts of MRI images and ultimately lead to better patient outcomes through more accurate diagnoses and treatments. Besides, the artifact extraction architecture of UNAEN can be generalized in other aspects of image quality improvement, such as reducing different types of artifacts, deblurring, and denoising. The possibility of these extensions will be further verified in our future work.

Despite the superior artifact reduction effect of UNAEN, we acknowledge some limitations. Firstly, we generated artifacts of brain MRI only through periodic motion patterns, while the movement of patients can be more complex and irregular in real MRI measurements. The performance of the proposed model trained with authentic MA-corrupted and MA-free images remains to be investigated. Besides, although UNAEN outperforms other state-of-the-art methods for MAR in terms of both quantitative metrics and visual quality, it is still an unsupervised learning-based method. It may not perform as well as supervised methods in the evaluation metrics when paired training data are used. Therefore, the network needs further optimization to mitigate the gap between the two types of different feature learning mechanisms.

\vspace{1cm}
\section{CONCLUSION}\label{sec conclusion}

In this paper, we proposed an improved GAN-based MRI motion artifact reduction network named UNAEN, which is unsupervisedly trained with unpaired MR images to circumvent the difficulty of obtaining paired MR images. UNAEN considers motion artifacts as the representable abnormality of MA-corrupted images and explicitly extracts and removes it without other transformations, which shows superiorities over the previously used image-to-image translation methods and effectively clears artifact components from MA-corrupted images. Our experimental results show that UNAEN alleviates the problem of lacking paired MA-corrupted and MA-free images and generates higher evaluation metrics and visual quality compared with some baseline models. Therefore, the proposed unpaired deep learning scheme has the potential to revolutionize clinical applications of MR imaging.

\bibliographystyle{IEEEtran}
\bibliography{reference}
\end{document}